\documentclass[aps,prc,twocolumn,superscriptaddress,showpacs,showkeys]{revtex4-1}
\usepackage{graphicx}

\begin{document}
\title{Search for neutrino oscillations at a research reactor}
\author{A.V.~Derbin}
\author{A.S.~Kayunov}
\author{V.V.~Muratova}
\affiliation{St.Petersburg Nuclear Physics Institute, Gatchina, Russia 188300}
\begin{abstract}
A project of search for oscillations of reactor neutrinos on a short, 5 - 15 meters, baseline with liquid scintillation position-sensitive detector
POSEIDON is described. The oscillations of the electron antineutrinos into sterile neutrinos will be searched at 100 MW research reactor with the
size of the active zone of about 0.5 m. The values of the oscillation parameters available for research are in the region $\delta m^{2}$ = (0.3 - 6)
eV$^{2}$ and Sin$^{2}$(2$\theta ) \geq $ 0.01.
\end{abstract}


\maketitle

\section{INTRODUCTION}

At present there is compelling evidence for the existence of sterile neutrino with the mass splitting $\delta m^{2} \quad \sim $ 1 eV$^{2}$ relative
to three known mass states and mixing parameter Sin$^{2}$(2$\theta ) \quad \sim $ 0.1. Since the decay width of the Z-boson indicates that the number
of light neutrino flavor states does not exceed three, a new sterile neutrino has no weak interaction. The discovery of the new fourth (or even
fifth) neutrinos would be beyond the scope of SM and would mean the existence of new physics unknown to us. The possible presence of one or two
sterile neutrino states points towards non-standard neutrino physics beyond the Standard Model (SM).

The schemes with a fourth massive neutrino with a mass of $\sim $ 1 eV$^{2}$ have attracted attention since the LSND group announced an excess of
electron antineutrinos in muon antineutrino beam \cite{Aguilar:2001}. This evidence has been confirmed by the MiniBoone experiment in the
antineutrino and the neutrino sector \cite{Aguilar:2010}. The data from the KARMEN detector exclude the region of large values $\delta $m$^{2}\geq$
($\sim $ 5 eV$^{2})$ \cite{Armbruster:2002}.

The other indications of short oscillations with $\delta m^{2} \quad \sim $ 1 eV$^{2}$ were obtained by radiochemical Ga-Ge-detectors SAGE and
GALLEX. The measurements with the artificial neutrino sources $^{51}$Cr and $^{37}$Ar \cite{Gavrin:1,Giunti:1} indicate a deficit of electron
neutrinos and it may imply neutrino oscillations.

Recently, the new calculations of the spectrum of reactor neutrinos lead to the value of the neutrino flux by about 3{\%} greater than which is
consistent with the experiments \cite{Mueller:2011,Mention:2011}. As a result, the data of the previous reactor experiments can be interpreted as the
observation of a flux deficit, and this, in turn, can be explained by additional sterile neutrinos with the masses at the eV scale

At last, the current data from cosmology and Big Bang nucleosynthesis favor an additional relativistic degree of freedom which can be interpreted as
sterile neutrino species \cite{Hamann:1}.

Some experiments to search for oscillations with $\delta m^{2} \quad \sim $ 1 eV$^{2}$ were proposed at reactors and at accelerators and with
artificial neutrino sources. The value $\delta m^{2} \quad \sim $ 1 eV$^{2 }$ corresponds to the ratio of oscillation length to neutrino energy
L/E$_{\nu }\approx$ 1 m/MeV = 1 km/GeV.

The Borexino collaboration plan to perform measurements with artificial sources of neutrino ($^{51}$Cr, $^{37}$Ar) and antineutrino ($^{144}$Ce,
$^{90}$Y+$^{90}$Sr) which makes it possible to check further CPT invariance \cite{Ranucci:2011,Pallavicini:1,Ianni:2011}. The Borexino detector
registers the scattering of neutrinos by electrons and the reaction of inverse beta-decay for antineutrinos \cite{sim:2011,sim:2012}.

The possibility of observing neutrino oscillations with $^{51}$Cr source, through the allocation of space zones of Ge production, is considered for
the detector SAGE \cite{Gorbachev:1}.

The experiment with the sources of $^{144}$Ce and $^{106}$Ru, which can be placed in the center of a large volume scintillation detector such as SNO,
BOREXINO or KamLAND is proposed in \cite{Cribier:1}. Detectors Nucifer (Saclay/IN2P3) \cite{Porta:2010} and DANSS (JINR / ITEP) \cite{Egorov:2011}
which are created to control the nuclear fuel, plan to use to search for oscillations at short distances. The experiments at accelerators with
L/E$_{\nu } \quad \sim $ 1 km/GeV are also considered \cite{workshop:2011}.

\section{ DETECTION OF ANTINEUTRINO}

The reaction of inverse beta decay $\nu  + p \to  e^{ + } + n$ is used for the registration of the antineutrino. Two consecutive events from the
positron and neutron capture in a short time window, which is determined by the lifetime of a neutron in the scintillator ($\sim $25 ms with 0.1{\%}
Gd mass), can reliably identify this reaction. The threshold of the reaction is 1.8 MeV and the positron energy is uniquely related to the
antineutrino energy. At a distance of 5 meters from 100 MW reactor the expected number of ($\nu p,n$e$^{ + })$ events is 3 / (kg day) for a target
containing 6x10$^{25}$ protons / kg (1 kg C$_{3n}$H$_{4n}$ scintillator).

The length of the oscillations for the two neutrinos mixing is: L [m] = 2.5 E$_{\nu}$ [MeV] / $\delta m^{2}$ [eV$^{2}$], and the depth of the
oscillations is determined by the angle of mixing Sin$^{2}$(2$\theta )$. The probability of observing the electron antineutrino ($\rm{P}_{ee})$
passing the path R is described by simple expression $\rm{P}_{ee}$ = 1 - Sin$^{2}$(2$\theta )$ Sin$^{2}(\rm{R/L})$. For a non-point source, as well
as for a non mono-energetic neutrinos, $\rm{P}_{ee}$ does not depend on $R$ for R $\gg$ L: $\rm{P}_{ee}$ = 1 - 0.5 Sin$^{2}$(2$\theta )$. In this
case, the oscillations appear through a decrease in the detected neutrino flux as compared with the flux calculated for the absence of oscillations.

\begin{figure}
\includegraphics[width=9cm,height=10.5cm]{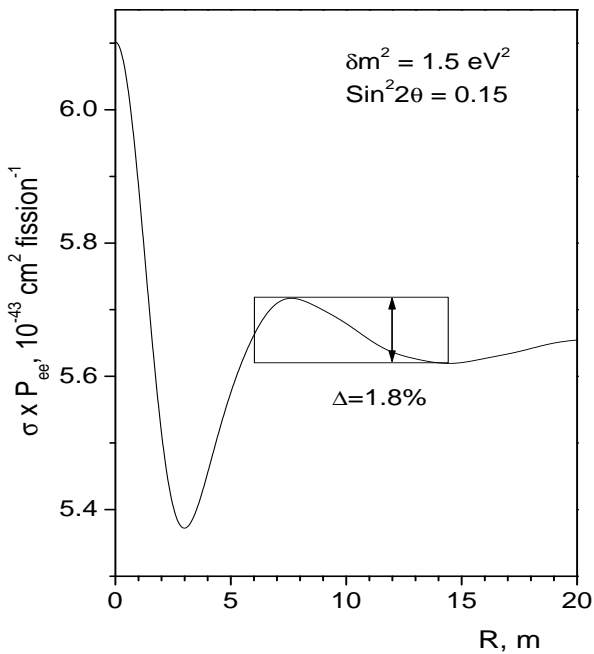}
\caption {The value of $\sigma (\nu $p, ne$^{ + })\times$P$_{ee}$ depending on the distance R for the neutrino detection threshold 1.8 MeV. }
\label{fig1}
\end{figure}

\begin{figure}
\includegraphics[width=9cm,height=10.5cm]{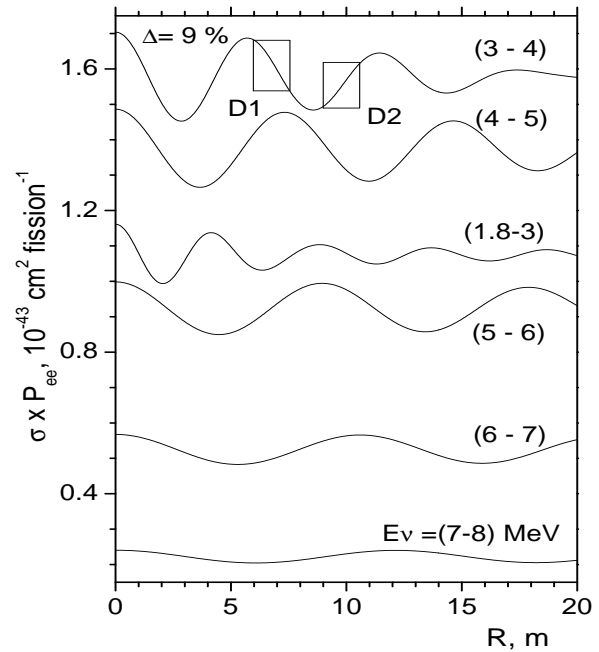}
\caption {The value of $\sigma (\nu $p, ne$^{ + })\times$P$_{ee}$ depending on R for the different ranges of the neutrino energy. } \label{fig2}
\end{figure}

We propose to look for the oscillation waves in the counting rate of antineutrinos of different energies associated with the transitions of the
electron antineutrino in a sterile one $\nu _{e} \quad  \to  \quad \nu _{s}$ and $\nu _{s} \quad  \to  \quad \nu _{e}$, directly on the size of the
detector. Two conditions are required to perform this. The size of the neutrino source must be smaller than the length of the oscillations. Since the
reactor neutrinos have a continuous spectrum, the energy resolution of the detector should be sufficient to observe changes in the shape of the
neutrino spectrum. The oscillation curves on the length of the detector are shown in Figures 1 and 2 for the values $\delta m^{2}$ = 1.5 eV$^{2}$ and
the angle of mixing Sin$^{2}$(2$\theta )$ = 0.15. These values are close to the ones obtained from the joint fit of the reactor, gallium an MiniBooNE
data ($\delta m^{2}\geq$ 1.5 eV$^{2}$, Sin$^{2}$(2$\theta )$ = 0.14 $\pm $ 0.08) \cite{Mention:2011}.

In the case of registration all neutrinos, with the energies above ($\nu p,n$e$^{ + })$ reaction threshold, the deviation from the law 1/$\rm{R}^2$
do not exceed 1.8{\%}, if the neutrinos are detected at a distance of more than 6 m (Fig. 1). But for the antineutrino with the energy in the range
(3 - 4) MeV, the change is 9{\%} over the length of the detector 1.5 m (Fig. 2). To observe these oscillations waves, the energy resolution of the
detector should be better than 0.5 MeV.

The most intense sources of neutrinos are nuclear power plants with thermal power of $\sim $ 3 GW, which provides the neutrino flux 10$^{13}$ cm$^{ -
2}$s$^{ - 1}$ at the distance of 20 meters. The characteristic size of the zone is about 5 meters that makes it impossible to observe the oscillation
waves if $\delta m^{2}\geq$ 1 eV$^{2}$. The zone of research reactors is much smaller and is about 0.5 meters. Thus, the zone of the high-flux
reactor PIK in Gatchina has the height of 50 cm and the diameter of 39 cm \cite{PNPI:2011}. The zone of the reactor SM-3 in Dimitrovgrad is even
smaller -- $42\times42\times35$ cm \cite{Reactors:2003}.

\section{ THE POSITION-SENSITIVE DETECTOR}

The proposed detector with the dimensions of central part $1.5\times1.0\times1.0$ m$^{3}$ is the liquid scintillator (LS) detector. The scintillation
signals are detected by 72 photomultipliers PMT125 with the flat photocathode 150 mm in diameter. PMTs are placed on two opposite yz-planes (if the
reactor active zone lies on x-axis) of the detector and they are separated from the central volume by 30 cm buffer. The buffer (no scintillation)
provides the spectrometric properties of the detector and protects the detector from radioactivity of PMTs. The central detector is surrounded by 4
additional $1.5\times1.2\times0.15$ m$^{3}$ detectors, each of which is viewed from the both sides by 14 PMTs. The sections perform the role of the
active shielding against cosmic rays and fast neutrons. They also detect 511 keV gammas which escape from the central detector after the positron
annihilation. The scintillators based on the linear alkylbenzene \cite{Barabanov:2011,Nemchenok:2011} (LAB, T = (130-150) C$^{o})$ or PXE
\cite{Back:2008} (C$_{12}$H$_{18}$, T = 167 C$^{o})$ are preferable because of a high flash point. The properties of PXE were studied in details on a
prototype Borexino detector \cite{Back:2008}. LAB is used by Daya Bay \cite{Wang:1} and Reno \cite{Soo:1} reactor experiments. Both solvents allow
the production of the Gd-containing scintillators.

The energy and coordinates of an event are reconstructed from the charge registered by PMTs. The Monte Carlo simulations based on GEANT4 were
performed to find the main parameters of the detector. Fig. 3 shows the detector response functions for 1 MeV (kinetic energy) positrons. The
positrons were uniformly simulated inside the central detector. A light yield equal to 10$^{4}$ photons/MeV is a real value for the scintillators
based on the LAB or PXE. For the quantum efficiency of the photocathode PMT125 equal to $\eta $ = 0.1, about 200 the photoelectrons (p.e.) at 1 MeV
energy release are registered, which provides a statistical component of the energy resolution of $\sigma $ = 7{\%} / E$^{1 / 2}$ where the energy E
is expressed in MeV units. The 511 keV escape peak is significantly suppressed for the case of additional sections of a thickness of 15 cm (Fig. 3,
curve 2).

\begin{figure}
\includegraphics[width=9cm,height=10.5cm]{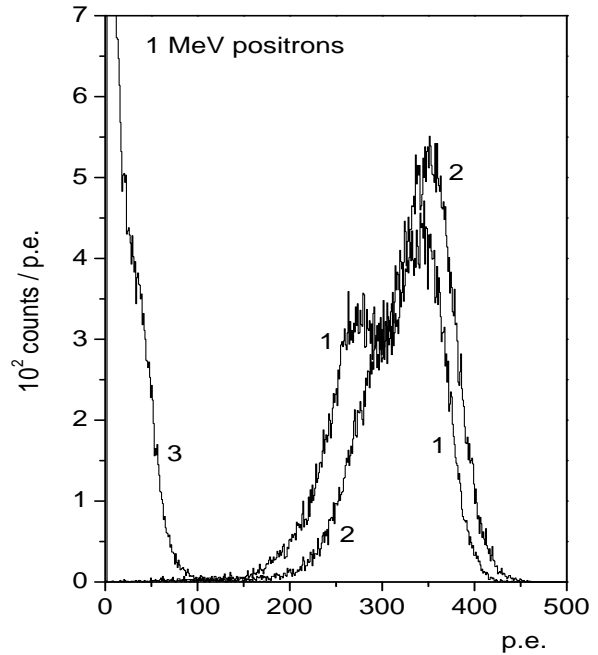}
\caption {The detector response functions for 1 MeV positrons: 1 - in the central volume ; 2 - in the whole volume, 3 - in 15 cm outer layer.
 } \label{fig3}
\end{figure}

\begin{figure}
\includegraphics[width=9cm,height=10.5cm]{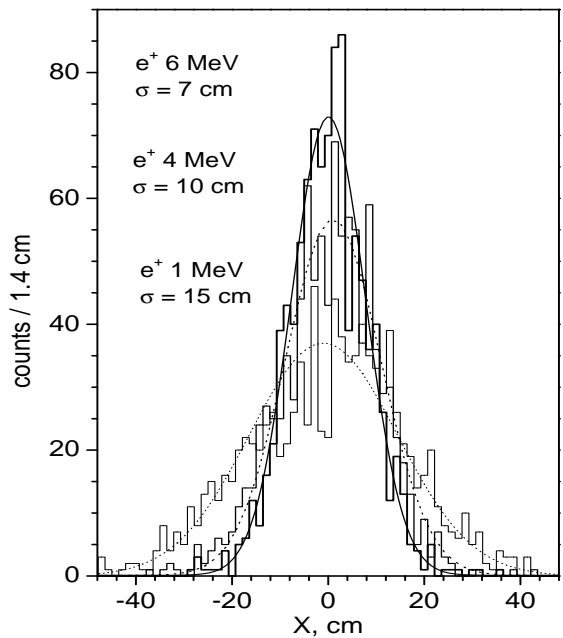}
\caption {The restore of x-coordinate for the events simulated in the center of the detector. } \label{fig4}
\end{figure}

To restore the coordinates of the event, the values of the charges (Q$_{i}$, i = 1 .. 72), registered with 72 PMTs are used. The reconstruction
program is based on a simple ratio of the charges collecting by PMTs from front and back sides of the detector. It provides a spatial resolution of
15 cm (1$\sigma )$ for the kinetic energy of positrons 1 MeV (Fig. 4). The reconstruction of the x-coordinate (in the direction of the reactor core)
with an accuracy of better than 15 cm allows observing oscillation waves on the length of the detector. At the same time this allows to improve the
energy resolution due to the space position and to compensate for the loss of the detection efficiency in the edges of the detector.

The spatial resolution for the neutron ($\sigma $ = 25 cm) is significantly worse than for the positron, however, it can be used as an additional
criterion for the selection of inverse $\beta $-decay events. In addition, time signals from PMTs allow us to distinguish the signals from electrons,
$\alpha $-particles and recoil protons due to different photons time responses.

The background level is the main problem of the experiment carried out at the Earth's surface and at a minimum distance from the reactor. The lead
shield with the thickness of 200 g/cm2 suppresses the external gamma-ray activity almost completely. The copper or iron shield is undesirable because
of the high-energy gamma rays from the capture of thermal neutrons. The hydrogen-consisting shield such as borated polyethylene reduces the flux of
fast and thermal neutrons inside the detector. All construction materials are tested for radioactive purity in the low-background setup with
HPGe-detector. The cosmic rays background is reduced by the active shield constructed from plastic scintillators. The relatively small size of the
main detector allows to use 2 layers of the active shield before and after the lead shield which is the generator of neutrons.

PMTs operate in a mode close to one-electron regime. The amplifier has a short restore time after pulses corresponding to a muon passing through the
LS. The count rate for a single PMT set about 100 Hz. The trigger appears if more than N PMTs fire in the time window of $\sim $ 20 ns. E.g. N = 20
corresponds to the energy threshold of 100 keV. The states of all PMTs are recorded in the time range 50 ms before and after the trigger. The charge
and time signals are used to reconstruct the position and energy of the event. For the trigger rate equal to 10 s$^{ - 1}$, about 1 GB of data are
stored per day. The neutrino events are selected in the off-line mode.

\section{ THE MAIN SOURCES OF THE BACKGROUND}

The main difficulty of this experiment is that it is carried out almost on the surface of the Earth. The suppression of the muons and the
nuclear-active component of cosmic rays is with only a few (less than 10) meters of water equivalent. The detector background is the main problem of
the experiment. A relatively small neutrino flux will be detected during reasonable time only if the background is comparable to the effect.

The main source of the background not correlated with the reactor is induced by random coincidences within a time window $\tau $ when signals from
positron and neutron are selected. The number of random coincidences is determined by the value N$_{1}$N$_{2}\tau $ where N$_{1}$ and N$_{2}$ are the
count rate of in the ranges where positrons and neutrons are registered. The value $\tau $ = 25 ms obtained for a 0.1{\%} Gd concentration can not be
reduced significantly. As a result, the value N$_{1}$N$_{2}$ must to be less than 300 s$^{ - 2}$, in order to the background of random coincidences
does not exceed the number of inverse beta-decays. To decrease N1N2 it is needed to increase the registration thresholds for positron and neutron or
to sectionalize the detector.

The background due to cosmic rays is suppressed by an active shield. Since the count rate of veto is at least 10$^{3}$ s$^{ - 1}$, the length of veto
can not exceed 100 ms to avoid a large dead time. As a result, the neutrons produced by cosmic rays can give a signal behind the veto pulse. Another
source of the background induced by long-lived light nuclei ($^{8}$He, $^{9}$Li, etc) which decay with neutron emission. The rate of the production
of such nuclei by muons in the Earth's surface is $\sim $ 150 events / (t day). The production of long-lived light nuclei by the nuclear-active
component will be measured with the detector prototype.

The natural radioactivity of the detector materials and the first layer of the passive shield determines the registration threshold for positrons.
The $^{40}$K in a glass PMT is the main source of the background. Due to $^{40}$K the threshold for positron will be set above 1.5 MeV that
corresponds to the minimum neutrino energy E$_{\nu} $ = 2.3 MeV. Due to the radioactivity of the lead shield and other inner components, the
threshold may be increased beyond the natural gamma activity of 2.6 MeV (E$_\nu $ = 3.4 MeV). The total energy of the gammas produced by neutron
capture by Gd nuclei is 8 MeV. The MC simulations show that the threshold of 2.6 MeV for the delayed signal corresponds to the detection efficiency
of 80{\%}.

To suppress the background of fast neutrons, correlated with the reactor power the difference in time response of the scintillator for electrons and
the recoil protons is used ($\beta /p$ discrimination). This discrimination suppresses the background of successive Bi-Po $\beta $ - and $\alpha
$-decay in uranium and thorium families.

\begin{figure}
\includegraphics[width=9cm,height=10.5cm]{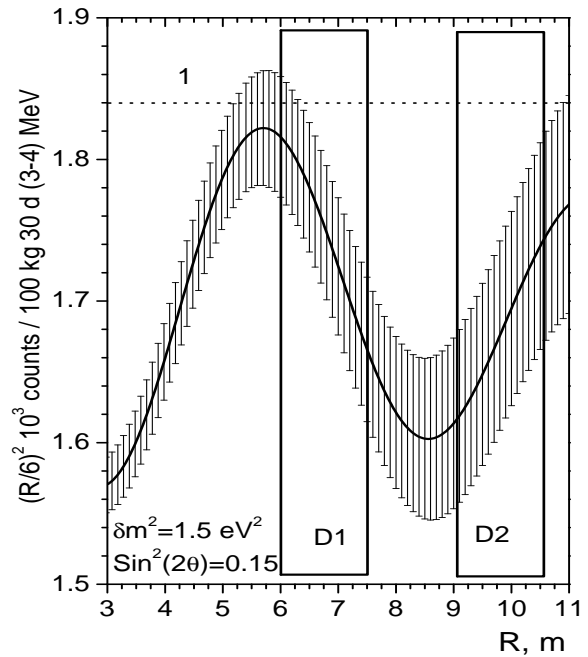}
\caption {The count rate multiplied by the value (R [m] / 6 m) $^{2}$. The errors correspond to 100 days of data taking. Dotted line is
no-oscillation case. } \label{fig5}
\end{figure}

\section{ SENSITIVITY OF EXPERIMENT}

We defined the sensitivity of the experiment for the neutrino energy range (3 - 4) MeV where the count rate is 900 events per day if the front wall
of the detector is located at a distance of 6 m from the reactor core. Fig. 5 shows the count rate of the reaction of inverse beta decay depending on
the distance from reactor core. The calculations take into account the finite size of the active zone and the energy and spatial resolution of the
detector. If Sin$^{2}$(2$\theta )$ = 0.15 the difference in the number of counts for front (R = (6-6.75) m) and back (R = (6.75-7.5) m) volume of the
detector is 4.5{\%}. By assuming that the background is equal to the effect and detection efficiency close to 100{\%} about 200 days of taking data
are needed to measure Sin$^{2}$(2$\theta )$ with an accuracy of 0.04. The number of events for a neutrino detected in the range 3-4 MeV is only 28
{\%} of the full neutrino events.

The proposed scheme of the experiment is realized only if the oscillation length lies in the range (1.5 - 30) m. The values of the parameters of
neutrino oscillations, which can be studied using the deviations from 1/R$^{2}$, lie in the region $\delta m^{2}$ = (0.3 - 6) eV$^{2}$ and
Sin$^{2}$(2$\theta )\geq$ 0.01. The sensitivity and reliability of the experiment increase by using two identical detectors D1 and D2 at two
distances of 6 and 9 meters (Fig. 6).

\section{ CONCLUSION}

The project with a liquid-scintillation position-sensitive detector (POSEIDON -position-sensitive detector of neutrino) to search for oscillations of
reactor neutrinos at short (5 - 15) meter base is proposed. The project aimed to search for oscillations of the electron antineutrinos into sterile
ones at 100 MW research reactor with 0.5 m active zone. The oscillation parameters which can be investigated are in the range $\delta m^{2}$ = (0.3 -
6) eV$^{2}$ and Sin$^{2}$(2$\theta)\geq$ 0.01.

\end{document}